\documentclass[aps,showpacs,preprintnumbers]{revtex4}
\usepackage{hyperref,amsmath,graphicx,subfigure}

\begin{document}%
\title{Gravitino cosmology in supersymmetric warm inflation} 
\date{\today}
\author{Sam Bartrum}
\email{s.bartrum@sms.ed.ac.uk}
\author{Arjun Berera}
\email{ab@ph.ed.ac.uk}
\author{Jo\~{a}o G.~Rosa}
\email{joao.rosa@ed.ac.uk}
\affiliation{SUPA, School of Physics and Astronomy, University of Edinburgh, Edinburgh, EH9 3JZ, United Kingdom}

\preprint{Edinburgh 2012/15}

\begin{abstract}

In supersymmetric models of warm inflation, the large temperature of the radiation bath produced by the dissipative motion of the inflaton field may induce a significant thermal abundance of potentially dangerous gravitinos. While previous discussions of this problem focused on gravitino production only at the end of warm inflation, similarly to conventional reheating scenarios, we study the full evolution of the gravitino abundance during and after inflation for simple monomial potentials, taking into account the enhanced gravitino and possibly gaugino masses due to supersymmetry breaking during inflation and the smooth transition into a radiation-dominated era. We find, on one hand, that the continuous thermal production increases the gravitino yield, although, on the other hand,  `freeze-out' occurs at temperatures much lower than previously estimated. Moreover, for sufficiently strong dissipation, which allows for sub-planckian inflaton values, the lower radiation temperature significantly alleviates and possibly solves the gravitino problem, with a baryon asymmetry being nevertheless produced through dissipative effects. Our analysis may also be relevant to standard reheating as an oscillating inflaton will also change the gravitino mass, potentially modifying the produced gravitino yield.

\end{abstract}
\maketitle


\section{Introduction}
Inflation \cite{Guth1981a,Albrecht:1982wi,Linde:1981mu} has been incredibly successful in providing solutions to the problems of the standard cosmological model. It can set the initial conditions, which give rise to the high degree of flatness and homogeneity that we observe in the universe today. From particle physics motivated models, it not only yields a mechanism for accelerated expansion but also explains, through quantum fluctuations, the origin of the temperature anisotropies in the Cosmic Microwave Background and the seeds for the observed Large Scale Structure.

In the standard cold or isentropic inflation scenario, the early universe is dominated by the vacuum energy of a scalar field which is slowly rolling down its potential, resulting in a period of accelerated expansion. This occurs whilst its kinetic energy is negligible compared to the potential energy until, at some point, the potential typically steepens and the inflaton begins oscillating about the minimum of its potential. During the period of accelerated expansion, the inflaton is assumed to have negligible couplings to other fields in order to keep the potential flat enough for a sufficient number of e-folds of inflation ($\sim$40-60) to occur and as a result the universe supercools. However, once it begins oscillating, there must be interactions that convert the vacuum energy into radiation in order to reheat the universe.

Hence, the inflaton cannot be an isolated system and, while the standard picture assumes that any interactions have a negligible effect on the dynamics of inflation and only become important during reheating, this need not be the case. In the alternative warm inflation paradigm, such interactions may in fact lead to dissipation of the inflaton's kinetic energy into light degrees of freedom, which in the simplest case may thermalise, resulting in the presence of a nearly-thermal bath concurrent with the accelerated expansion.

In the early universe, gravitinos can be abundantly produced, potentially leading to overclosure of the universe or spoiling the abundances of light elements predicted by the standard big bang nucleosynthesis model (BBN). While in cold inflation thermal production of gravitinos occurs only during the reheating phase, in warm inflation this is concurrent with inflation due to the presence of a thermal bath. In standard reheating, gravitino overproduction constrains the reheat temperature, i.e. the maximum temperature after inflation when the universe becomes radiation dominated. There is, however, a certain amount of tension in this case between having a large enough reheat temperature to allow for a thermal mechanism for baryogenesis, whilst keeping it low enough to avoid overproducing gravitinos. In warm inflation, on the other hand, this tension can be relieved, as a baryon asymmetry may in fact be produced at low temperatures through dissipative effects \cite{Bastero-Gil2011}, potentially avoiding overproduction of gravitinos.

Gravitino production in warm inflation has been considered previously in \cite{Taylor2001,Bastero-gila}, where it was assumed that the effective `reheat temperature' occurs when the radiation energy density becomes equal to the inflaton energy density, $\rho_{\phi}=\rho_R$, and that standard reheating constraints on gravitino production can be applied. This may, however, overestimate the temperature at which the gravitino yield freezes out, as radiation does not yet fully dominate the energy density at this stage. Moreover, standard reheating constraints may not {\it a priori} be applied in warm inflation scenarios due to the non-negligible abundance of gravitinos produced during inflation, which may potentially lead to a larger yield. This may, in fact, be the case also in conventional models, since the reheating phase is not necessarily instantaneous and thermal production of gravitinos may potentially occur for the duration of reheating and not freeze out until the universe is fully radiation dominated, resulting in a cumulative effect similar to that of warm inflation. Finally, we note that supersymmetry is broken during inflation, leading to gravitino masses parametrically close to the Hubble parameter and potentially to massive gauginos, which may also modify the production rate during warm inflation. Similarly, this may change the standard reheating constraints, as the gravitino mass also varies during the oscillating phase.


With these new insights in mind, we revisit the production of gravitinos in supersymmetric warm inflation, numerically evolving the Boltzmann equation for gravitinos into the radiation era. In Section \ref{SGC} we give a brief review of the standard gravitino cosmology in cold inflation and in Section \ref{WI} we outline the basic features of warm inflation, focusing as a working example on  monomial potentials in the sub-planckian regime. We discuss thermal gravitino production in warm inflation in Section \ref{Gravprod} and present results for stable and unstable gravitinos, considering the effects of inflaton-dependent gaugino masses in both cases. In Section \ref{Discussion} we summarize our main results and discuss possible directions of future research in this topic.


\section{Standard Gravitino Cosmology}\label{SGC}

Supersymmetry is an attractive theory for inflationary dynamics due to the presence of a whole host of scalar fields, as for example the superpartners of Standard Model quarks and leptons, automatically protecting the scalar potential from quadratic loop corrections that may spoil its required flatness. However, several single field models typically require inflaton expectation values close to the Planck scale, where supergravity effects start playing an important role. The gauge particle of supergravity is the massless spin-$3/2$ gravitino and, when supersymmetry is broken, the gravitino becomes massive and absorbs the spin-$1/2$ goldstino through the super-Higgs mechanism. Due to its indiscriminate coupling, the neutral gravitino couples to all fields universally, whether in Standard Model/visible sector or other hidden/sequestered sectors, with planck-suppressed interactions, making it a potential candidate for dark matter. This suppressed coupling makes it, however, unlikely to be detected at man-made colliders. Gravitinos may nevertheless be abundantly produced in the early universe through a variety of thermal and non-thermal processes due to the large energies involved. Unfortunately, due to our ignorance of the mechanism behind supersymmetry breaking, its mass is unknown and can only be constrained by cosmological considerations. 

Without a period of inflation the constraints on the gravitino mass are quite severe. In the standard cosmological model, the early universe is radiation-dominated and, at early enough times, the temperature will be high enough for gravitinos to be in thermal equilibrium with the radiation bath. For stable gravitinos, the early freeze-out associated with planck-suppressed interactions can thus result in overclosure of the universe unless $m_{\tilde{G}}\lesssim1\,$keV \cite{Pagels1982}. For an unstable gravitino, its mass needs to be larger than $\sim10\,$TeV, otherwise it will decay during BBN and spoil the predictions for the light element abundances \cite{Weinberg1982a}. With a period of inflation, any initial population of gravitinos is diluted away and, in cold inflation models, no gravitinos are thermally produced until reheating. These strict bounds on the gravitino mass are thus somewhat relaxed and replaced by upper limits on the reheat temperature \cite{Khlopov:1984pf}.

Gravitinos are primarily produced by the scattering of particles in a thermal bath. Due to the stronger coupling, the dominant production comes from inelastic $2\rightarrow 2$ QCD processes involving left handed quarks ($q$), squarks ($\tilde{q}$), gluons ($g$), gluinos ($\tilde{g}$) and gravitinos ($\tilde{G}$) such as $g+g\rightarrow \tilde{g}+\tilde{G}$, $q+\bar{q}\rightarrow \tilde{g}+\tilde{G}$ and $\tilde{q}+g\rightarrow \tilde{q}+\tilde{G}$. While this contribution to the thermal gravitino production rate in supersymmetric QCD has been calculated in \cite{Bolz2000}, in this work we will adopt the complete $SU(3)_c\times SU(2)_L\times U(1)_Y$ thermal production rate computed in \cite{Pradler2007a}. Gravitinos can also be produced from the decay of the inflaton during its oscillating phase after inflation (see \cite{Kawasaki2006,Hashimoto1998,Endo2006,Kohri2004,Endo2007,Fujisaki,Nilles2001}). However, as we discuss below, this only becomes significant during the radiation era and, due to the suppression of inflaton oscillations for strong dissipation, it is subdominant for the monomial models we consider. They can also be produced through the decay of other particles in the thermal bath, assuming this is kinematically allowed. However, during inflation the Hubble parameter is in general much larger than the relevant planck-suppressed decay widths, $H\gg \Gamma_{decay}$, so that these decays will also not occur until the Hubble parameter drops significantly in the radiation era. As it is our aim to highlight the differences between the gravitino production in warm and cold inflation, we will thus focus on thermal production processes.

If gravitinos are abundantly produced in the early universe, this can pose a problem for inflationary model building. 
There are two situations to consider in gravitino cosmology, depending on whether the gravitino is the lightest supersymmetric partner (LSP) and stable by virtue of R-Parity conservation, or otherwise the gravitino is unstable and will decay at some stage in the cosmological evolution. Constraints on primordial abundances come in either case from two sources. Firstly, the abundance of the LSP must not exceed the observed dark matter abundance (\ref{DMconstraint}) \cite{Yao:2006px}: 
\begin{equation}\label{DMconstraint}
 \Omega_{DM}h^2=0.105^{+0.007}_{-0.010}~.
\end{equation}
Secondly, the decay products of the next to LSP (NLSP) must not spoil BBN predictions for light element abundances. Radiative decay of the NLSP where photons and charged particles are emitted can induce electromagnetic showers, disintegrating the light elements. The NLSP can also decay into quarks or gluons, which subsequently hadronise. These hadrons can then induce interconversions between the background protons and neutrons, enhancing the neutron to proton ratio and thus resulting in an overproduction of $^4$He. The energetic nucleons can also destroy the background $^4$He and non-thermally produce D, T, $^3$He, $^6$Li and $^7$Be. There is also the possibility that if the LSP is charged then it could bind with background nuclei and change the nuclear reaction rates, in particlar that of $^6$Li.

If the gravitino is unstable and $m_{\tilde{G}}\lesssim 20\,$TeV, its lifetime is longer than 1 s \cite{Kawasaki2008} and so it will be subject to the BBN constraints mentioned above. It will also decay into the LSP which needs to satisfy the dark matter constraint in Eq.~(\ref{DMconstraint}). If the gravitino is stable then it must satisfy the dark matter constraint (\ref{DMconstraint}) and the decay of the NLSP into the gravitino must avoid upsetting BBN predictions. For more details on BBN constraints on LSP and non-LSP gravitino primordial abundances see \cite{Falomkin:1984eu,Kawasaki2008,Feng2004a,Feng2004,Kawasaki1995,Kawasaki2005a,Kohri2006,Kawasaki:2004qu,Pradler2006}. The problems mentioned above constitute the so called ``gravitino problem''.

The number density of gravitinos is described by the Boltzmann equation:
\begin{equation}\label{BoltzGrav}
 \dot{n}_{\tilde{G}}+3Hn_{\tilde{G}} = C_{\tilde{G}}~,
\end{equation}
where we neglect gravitino decay and gravitinos produced from decays of other fields. The collision term, $C_{\tilde{G}}$, describes gravitino production in a thermal bath and is given by \cite{Pradler2007a}:
\begin{equation}\label{Collision}
 C_{\tilde{G}}=\frac{3\zeta(3)T^6}{16\pi^3m_p^2}\sum_{i=1}^3\left(1+\frac{m_{\tilde{g}_i}^2}{3m_{\tilde{G}}^2}\right)c_ig_i^2\log\left(\frac{k_i}{g_i}\right)~.
\end{equation}
The index $i$ runs over the gauge groups $(U(1)_Y,SU(2)_L,SU(3)_c)$ where $m_{\tilde{g}_i}$ are the gaugino masses, $g_i$ are the gauge couplings and $c_i=(11,27,72)$, $k_i=(1.266,1.312,1.271)$.
The reheating phase is assumed to be instantaneous, immediately entering the radition era. 
Defining the gravitino-to-photon yield, $Y_{\tilde{G}}=n_{\tilde{G}}/n_{\gamma}$, with $n_\gamma=2\zeta(3)T^3/\pi^2$, and assuming that $TR=$ constant, where $R$ is the scale factor, we obtain from Eq.~(\ref{BoltzGrav}):
\begin{equation}\label{yieldevol}
 \frac{dY_{\tilde{G}}}{dT}=-\frac{C_{\tilde{G}}}{H(T)T n_{\gamma}(T)}~.
\end{equation}
As $C_{\tilde{G}}\sim T^6$, with only a mild temperature dependence from the couplings and gaugino masses, Eq.~(\ref{yieldevol}) can be approximately integrated. Assuming that any initial population of gravitinos before the reheating phase is diluted away, $Y_{\tilde{G}}(T_R)=0$ and that we are interested in the yield of gravitinos at temperatures $T\ll T_R$, e.g.~at BBN, then:
\begin{equation}\label{gravyield}
Y_{\tilde{G}}(T)\approx\frac{C_{\tilde{G}}(T_R)}{H(T_R)n_{\gamma}(T_R)}~.
\end{equation}
However, $TR$ only remains constant away from particle mass thresholds, as it is instead the entropy density that is conserved,  $sR^3=$constant. We can take this into account by diluting the yield:
\begin{eqnarray}
 Y_{\tilde{G}}(T_1)=\frac{s(T_1)/n_{\gamma}(T_1)}{s(T_2)/n_{\gamma}(T_2)}Y_{\tilde{G}}(T_2)
=\frac{g_*(T_1)}{g_*(T_2)}Y_{\tilde{G}}(T_2)~,
\end{eqnarray}
where $g_*(T)$ is the number of relativistic degrees of freedom in thermal equilibrium at temperature $T$. Note that this is generically well below the gravitino yield in thermal equilibrium. We then obtain the following expression for the present abundance of gravitinos:
\begin{eqnarray}\label{Steffan}
\Omega_{\tilde{G}}^{th}h^2 &=&m_{\tilde{G}}n_{\tilde{G}_0}\rho_c^{-1}h^2\nonumber \\
&=&\sum_{i=1}^3\omega_ig_i^2\left(1+\frac{M_i^2}{3m_{\tilde{G}}^2}\right)\ln\left(\frac{k_i}{g_i}\right)\left(\frac{m_{\tilde{G}}}{100GeV}\right)\left(\frac{T_R}{10^{10}GeV}\right)~.
\end{eqnarray}
The subscript `$0$' indicates the present day value, with $T_0=2.73\,$K, $g_*(T_0)=3.91$. We use the MSSM value for $g_*(T_R)=228.75$ and the critical density $\rho_c=8.1\times10^{-47}h^2\,$GeV$^4$ with the constants $\omega_i=(0.018,0.044,0.117)$. It is understood that the couplings and masses should be evolved with the temperature. This provides the standard cold inflation constraints on the reheat temperature and gravitino mass to avoid overclosure for an LSP gravitino. It is evident that for $m_{\tilde{G}}\approx100\,$GeV, to avoid overclosure, the reheat temperature $T_R\lesssim10^{10}\,$GeV. If the gravitino is the NLSP, then each gravitino will decay into one LSP and the primordial gravitino yield in Eq.~(\ref{gravyield}) can be converted into the LSP yield through:
\begin{equation}\label{LSPconvert}
 \Omega_{LSP}h^2 = \frac{m_{LSP}}{m_{\tilde{G}}}\Omega^{th}_{\tilde{G}}h^2~.
\end{equation}
We can see that $\Omega^{th}_{\tilde{G}}h^2>0.105$ is allowed as long as $m_{LSP}/m_{\tilde{G}}$ is sufficiently small. For unstable gravitinos with $m_{\tilde{G}}\lesssim20\,\,$TeV the strongest constraints come from BBN abundances, whilst above this the dark matter constraint for the LSP dominates \cite{Kawasaki2008}. These constraints are often defined in the literature in terms of the gravitino-to-entropy yield, $Y_{\tilde{G}}^s=n_{\tilde{G}}/s$, which can be easily related to the more convenient definition in terms of the photon energy density used in Eq.~(\ref{yieldevol}). We will take the conservative bounds of $Y_{\tilde{G}}^s\lesssim10^{-16}$ for $100\,$GeV$\lesssim m_{\tilde{G}}\lesssim1\,$TeV and $Y_{\tilde{G}}^s\lesssim10^{-14}$, $Y_{\tilde{G}}^s\lesssim10^{-17}$ for $1\,$TeV$\lesssim m_{\tilde{G}}\lesssim3\,$TeV for branching ratios into hadrons of $B_h=10^{-3}$, $B_h=1$ respectively \cite{Kawasaki2005a}. 
For stable gravitinos the BBN constraints on the primordial yield
from NLSP decays are quite model dependent, varying upon which particle is the NLSP, as well as its thermal abundance and mass. For more details on scenarios with sneutrino, slepton and neutralino NLSPs, see e.g.~\cite{Feng2004a,Feng2004}.


\section{Warm Inflation}\label{WI}

Warm inflation \cite{Berera1995b, Berera1995a, Berera:1996nv} (see also \cite{Fang:1980wi, Moss:1985wn, Yokoyama:1987an}) is an alternative picture of inflation, where the inflaton has non-negligible interactions with other fields that lead to fluctuation-dissipation dynamics and associated particle production concurrent with accelerated expansion. While several efforts in the literature to analyze non-equilibrium dynamics fall within this category (see e.g. \cite{Barnaby:2009dd, LopezNacir:2011kk, Barnaby:2012tk, Barnaby:2012xt}), the most well understood and extensively studied scenarios in the context of quantum field theory consider the case where radiation is produced in a nearly-thermalized state \cite{Berera:1996nv, Berera:1998gx, Berera:1998px, Berera:1999ws, Berera:2002sp, Berera:2008ar}. This allows one to make accurate predictions for the effects of dissipation and has several attractive features from the model-building point of view. Firstly, the dissipative dynamics acts as an additional source of damping that allows for longer periods of accelerated expansion, which is particularly important in supergravity/string theories, where F-term supersymmetry breaking typically induces large inflaton masses, thus alleviating the associated eta-problem \cite{Berera:1999wt,Berera:2004vm, BasteroGil:2009gh, Cai:2010wt, BasteroGil:2011mr}. Secondly, whilst one assumes radiation to be a subdominant component of the energy balance in the universe for accelerated expansion to occur, in several scenarios it may actually come to dominate at a later stage, providing a smooth transition into a radiation-dominated era, in alternative to the standard reheating picture. Finally, since thermal fluctuations overcome the quantum vacuum fluctuations for temperatures $T>H$, the spectrum of primordial density fluctuations may be significantly modified \cite{Berera1995a, Berera:1999ws, Taylor:2000ze, Hall:2003zp, Moss:2008yb}, in particular suppressing the amplitude of tensor perturbations and inducing potentially observable deviations from a gaussian spectrum \cite{Gupta:2002kn, Moss:2007cv, Chen:2007gd, Moss:2011qc}.  

The dissipative dynamics arising from interactions of the inflaton with other fields arises through time non-local contributions to its quantum effective action, which for a slow-rolling inflaton may in general be computed using linear response theory. In the adiabatic regime where $\dot{\phi}/\phi< \tau^{-1}$, where $\tau$ is the typical relaxation time of the nearly-thermal ensemble, this yields an effective friction term $\Upsilon\dot\phi$ in the equations of motion, which can be written as:   
\begin{equation}\label{IE}
\ddot{\phi}+3H(1+Q)\dot{\phi}+V_{\phi}=0~,
\end{equation}
where $Q=\Upsilon/3H$ and $V_{\phi}$ denotes the derivative of the potential with respect to the inflaton field.  Noting that the effective density and pressure of the inflaton condensate are given by $\rho_{\phi}= \dot{\phi}^2+V(\phi)$ and $p_{\phi}= \dot{\phi}^2-V(\phi)$, respectively, this can be rewritten as: 
\begin{equation}\label{ied}
\dot{\rho_{\phi}} + 3H(p_{\phi}+\rho_{\phi}) = -\Upsilon\dot\phi^2~.
\end{equation}
The energy lost by the inflaton field through dissipative effects is then gained by the produced particles (see e.g. \cite{Graham:2008vu}), and for $g_*$ relativistic degrees of freedom this yields the following evolution equation for the radiation density, $\rho_R=\pi^2g_*T^4/30$:
\begin{equation}\label{red}
\dot{\rho_R}+4H\rho_R = \Upsilon \dot{\phi}^2~,
\end{equation}
with inflation occuring for $\rho_{\phi}\gg\rho_R$. This nevertheless allows for $T>H$, as mentioned above, in which case one may also neglect the quasi-de Sitter expansion when computing the dissipation coefficient in different quantum field theory models. On the other hand, for $T<H$ we expect the inflationary dynamics to be similar to the more conventional cold scenarios.

Accelerated expansion occurs in the slow-roll regime, where $V(\phi)\gg\dot{\phi}^2$, $\ddot{\phi}\ll H\dot{\phi}$. In warm inflation, this can be translated into the modified slow-roll conditions:
\begin{equation}
\epsilon_{\phi}=\frac{m_p^2}{2}\left(\frac{V_{\phi}}{V}\right)^2<1+Q~, \hspace{1cm} \eta_{\phi}= m_p^2\left(\frac{V_{\phi\phi}}{V}\right)< 1+Q~,  \hspace{1cm} \sigma_\phi=m_p^2\frac{V_\phi}{V\phi}<1+Q~.
\end{equation}
In addition, we also require three more conditions. Firstly, we need the variation of $\Upsilon$ with respect to $\phi$ to be sufficiently slow, in order to avoid dissipation increasing too quickly and radiation dominating too soon. Secondly, we require that radiation is produced faster than it is diluted by the expansion of the Universe and, finally, that both quantum and thermal corrections to the inflaton potential are not too large and, in particular, do not induce a large inflaton mass. Once these conditions are violated either the radiation energy density starts to dominate or the inflaton is no longer overdamped and slow-roll ends (see e.g.~\cite{BasteroGil:2009ec}). In the slow-roll regime the equations of motion reduce to:
\begin{eqnarray}\label{sreom}
3H(1+Q)\dot{\phi}&\approx&-V_{\phi}~, \nonumber \\
4\rho_R&\approx&3Q\dot{\phi}^2~.
\end{eqnarray}

Earlier attempts to construct models of warm inflation considered a direct coupling between the inflaton and the light fields that form the radiation bath, but in this case a sufficiently large dissipation coefficient also induces a large thermal mass to the inflaton field, which makes it difficult to achieve a sufficiently long period of accelerated expansion \cite{Berera:1998gx, Yokoyama:1998ju}. A more promising avenue considers a two-stage mechanism \cite{Berera:2002sp}, where the inflaton is coupled to heavy fields that may subsequently decay into light degrees of freedom. This is also a more natural approach since couplings to the inflaton generically induce large masses. Moreover, in supersymmetric models the leading corrections to the inflaton potential are logarithmic in this regime \cite{Hall:2004zr}, despite supersymmetry being broken by the finite temperature and energy density, keeping the flatness of the potential stable against quantum and thermal corrections. A generic superpotential implementing this mechanism is given by \cite{Moss2006, BasteroGil:2009ec}: 
\begin{equation}\label{twostage}
W = W(\Phi)+g\Phi X^2+hXY^2~.
\end{equation}
The scalar component of $\Phi$ is the inflaton field, with expectation value $\phi=\varphi/\sqrt{2}$, which we assume to be real. Both the bosonic and fermionic components of the superfield $X$ then acquire masses proportional to $\varphi$ and can decay into the $Y$ scalars and fermions, which remain light and form the radiation bath. For $T\ll m_X$ and a broad range of couplings and field multiplicities, the leading contribution to the time non-local effective action corresponds to 1-loop diagrams involving virtual $X$-scalars, and has been discussed in \cite{Moss2006, BasteroGil:2010pb, Rosa2012}, yielding a dissipation coefficient of the form:
\begin{equation} \label{Upsilon}
\Upsilon\approx C_{\phi}\frac{T^3}{\phi^2}~,
\end{equation}
where $C_\phi$ depends on the coupling $h$ and the field multiplicities in the $X$ and $Y$ sectors. We restrict our analysis to a dissipative coefficient of the form (\ref{Upsilon}). Although our analysis depends upon the form of the dissipative coefficient we expect our methodology to be applicable to other forms \cite{Berera:1998gx, Yokoyama:1998ju, Rosa2012} and our qualitative results to be similar. In this work, we will take $C_{\phi}$ as a free parameter of the model, bearing in mind that large values for this constant may require a somewhat large field multiplicity, which may be attained, for example, in GUT models with large representations or the multiple D-brane constructions described in \cite{BasteroGil:2011mr}.


\subsection{Monomial potentials}

As a working example, we will take the inflaton superpotential to be of the form:
\begin{equation}\label{spinf}
W(\Phi) = \frac{\lambda}{r+1}\frac{\Phi^{r+1}}{m_p^{r-2}}~.
\end{equation}
For $r=0$, $\lambda<0$ we recover supersymmetric hybrid inflation, with the $X$ scalars corresponding to the waterfall field(s), and for $r>1$ we recover chaotic inflation models. Assuming a canonical K\"ahler potential for the inflaton field, $K(\Phi,\Phi^\dagger)=\Phi^\dagger\Phi$, this results in the following scalar potential:
\begin{equation} \label{scalar_potential}
 V = \lambda^2 m_p^4\left(\frac{|\phi|}{m_p}\right)^{2r}\left[1+\frac{1}{r+1}\left(\frac{|\phi|}{m_p}\right)^2\left(2-\frac{3}{r+1}\right)+\frac{1}{(r+1)^2}\left(\frac{|\phi|}{m_p}\right)^4\right]
\exp\left(\frac{|\phi|^2}{m_p^2}\right)~.
\end{equation}

Given our ignorance of fundamental quantum gravity effects, we will restrict our analysis to the sub-planckian regime $|\phi|\ll m_p$, where supergravity effects may also be ignored and the potential takes the simpler monomial form:
\begin{equation}
 V\approx\lambda^2\left(\frac{|\phi|}{m_p}\right)^{2r}m_p^4~.
\end{equation}
The slow-roll parameters are given by:
\begin{equation}
\eta_\phi = 2r(2r-1)\left(\frac{\phi}{m_p}\right)^{-2}~,\hspace{1cm} \epsilon_\phi=\frac{r}{2r-1}\eta~,\hspace{1cm}  \sigma_\phi = m_p^2\left(\frac{V_{\phi}}{\phi V}\right) = \frac{\eta}{2r-1}~.
\end{equation}
From the slow-roll equations of motion (\ref{sreom}), we can derive the following relation between $Q$ and $\phi$:
\begin{equation}\label{Qphirel}
 Q^{1/3}(1+Q)^2 = 2\epsilon_{\phi}\left(\frac{C_{\phi}}{3}\right)^{1/3}\left(\frac{C_{\phi}}{4C_R}\right)\left(\frac{H}{\phi}\right)^{8/3}\left(\frac{m_p}{H}\right)^2~,
\end{equation}
where $C_R=g_*\pi^2/30$. We can invert this to get:
\begin{equation}\label{srphi}
 \left(\frac{\phi}{m_p}\right)=\left(\frac{r^6C_{\phi}^4\lambda^2}{9C_R^3}\frac{1}{Q(1+Q)^6}\right)^{s}~,
\end{equation}
where $s=1/(14-2r)$. The evolution of $Q$ during inflation is found by differentiating Eq.~(\ref{Qphirel}) with respect to the number of e-folds.
\begin{eqnarray}\label{Qevol}
\frac{dQ}{dN_e}&=&\frac{Q}{1+7Q}(10\epsilon_{\phi}-6\eta_{\phi}+8\sigma_{\phi}) \nonumber \\
&=&\frac{Q^{1+2s}(1+Q)^{12s}}{1+7Q}\left[\frac{2r}{s}\left(\frac{9C_R^3}{r^6C_{\phi}^4\lambda^2}\right)^{2s}\right]~.
\end{eqnarray}
It is clear from Eq.~(\ref{Qevol}) that for $0<r<7$, $Q$ increases during inflation.
The number of e-folds of inflation can then be obtained by integrating Eq.~(\ref{Qevol}), giving:
\begin{eqnarray}\label{Neint}
 N_e = \int_{Q*}^{Q_e}\frac{dN_e}{dQ}dQ=C_Q \int_{Q_*}^{Q_e}\frac{1+7Q}{Q^{1+2s}(1+Q)^{12s}}dQ~,
\end{eqnarray}
with
\begin{equation}
 C_Q = \frac{s}{2r}\left(\frac{r^6C_{\phi}^4\lambda^2}{9C_R^3}\right)^{2s}~.
\end{equation}
The `$e$' subscript denotes the number of e-folds at which the slow-roll conditions are violated, while the `$*$' subscript indicates the value when cosmological scales leave the horizon during inflation. Performing the integral, this yields:
\begin{equation}
 N_e=C_Q[F_r(Q_e)-F_r(Q_*)]~,
\end{equation}
where
\begin{equation}
 F_r(x)=\frac{1}{2}x^{-2r}\left(\frac{14x}{1-2r}{_2}F_1(1-2x,12x,2-2x,-x)-\frac{1}{2} {_2}F_1(-2x,12x,1-2x,-x)\right)
\end{equation}
and $_2F_1(a,b,c,z)$ is the hypergeometric function. As $Q\rightarrow\infty$, the number of e-folds approaches a constant and so
it is not always possible to achieve the desired number of e-folds of inflation in areas of parameter space where $Q$ diverges too early, corresponding to the breakdown of the slow-roll approximation.

It is important to ensure that $T>H$, so that the dissipative coefficient in Eq.~(\ref{Upsilon}) can be calculated neglecting expansion effects. If we set $T_*/H_*>1$ at horizon crossing, then this will hold for the duration of inflation, for $0<r<7$, 
as we can see from Eqs.~(\ref{srphi}) and (\ref{srtemp}):
\begin{equation}\label{srtemp}
\frac{T}{H}=\left(\frac{9Q}{C_{\phi}\lambda^2}\right)^{1/3}\left(\frac{\phi}{m_p}\right)^{2(1-r)/3}~.
\end{equation}
As mentioned earlier, choosing $C_{\phi}$ as our other free parameter, the slow-roll dynamics are fully determined. We can use the amplitude of the primordial power spectrum,  $P_R^{1/2}\approx 5\times 10^{-5}$ \cite{Komatsu2009}, to fix $Q_*$:
\begin{equation}\label{ps}
P_{\mathcal{R}}^{1/2}\approx \left(\frac{H_*}{2\pi}\right)\left(\frac{3H_*^2}{V_{\phi}}\right)(1+Q_*)^{5/4}\left(\frac{T_*}{H_*}\right)^{1/2}~.
\end{equation}
Combining Eq.~(\ref{ps}) with the form of the dissipation coefficient in Eq.~(\ref{Upsilon}) and the relation between $Q$ and $\phi$ in Eq.~(\ref{Qphirel}), we arrive at:
\begin{equation}
(1+Q_*)^{1/2}Q_* = \frac{16\pi^2}{3}P_{\mathcal{R}}C_R\left(\frac{T_*}{H_*}\right)^{3}~.
\end{equation}
Note that $Q_*$ only depends on $(T_*/H_*)$ and not on $C_{\phi}$ or the form of the potential. Once we have $Q_*$, we can integrate Eq.~(\ref{Neint}) to obtain the total number of e-folds. The regime where $|\phi|\ll m_p$ corresponds to the strong dissipation limit, $Q_*\gg 1$. Making this approximation, we have:
\begin{equation}
 N_e = \frac{7}{2rs}\left(\frac{1}{Q_*^{2rs}}-\frac{1}{Q_e^{2rs}}\right)C_Q~.
\end{equation}

\begin{figure}[ht]
\centering
\subfigure{
\includegraphics[scale=0.81]{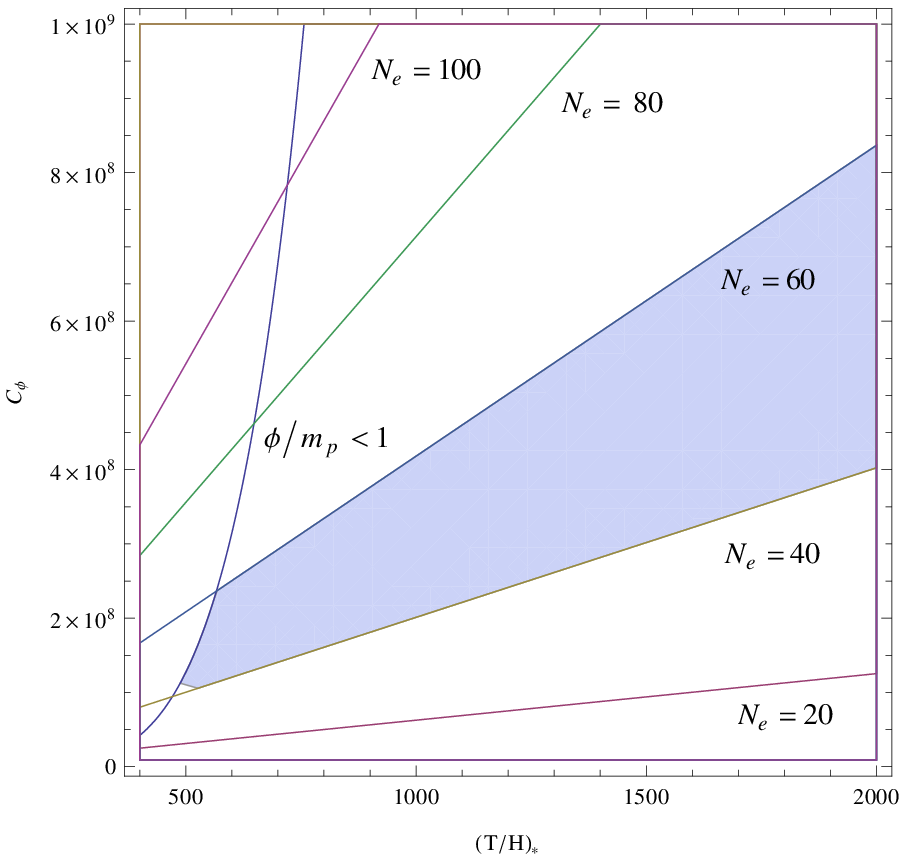}
  
}
\subfigure{
\includegraphics[scale=0.85]{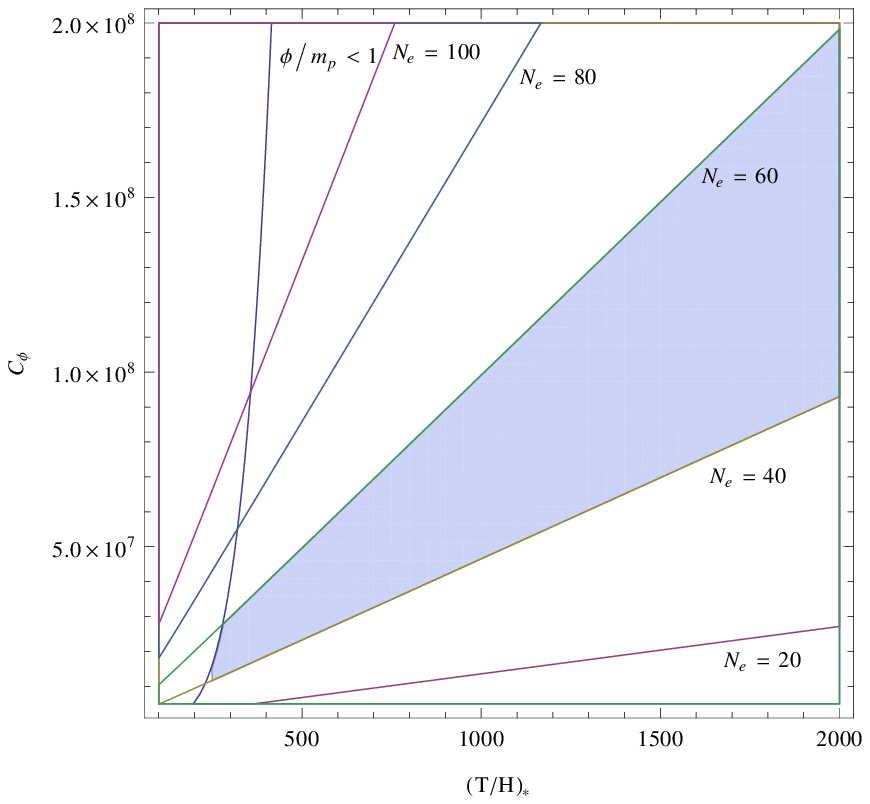}
   
}
\caption[Optional caption for list of figures]{Total number of e-folds for the quartic ($r=2$, left) and quadratic ($r=1$, right) potentials. The region where $|\phi|< m_p$ is to the right of the labelled line. The shaded region corresponds to between 40 and 60 e-folds of sub-planckian inflation.}
\label{Paraplots}
\end{figure}

In Figure \ref{Paraplots}, we show the region of parameter space for the quartic ($r=2$) and quadratic ($r=1$) potentials where we can ignore supergravity corrections with a reasonable number of e-folds of inflation. We note that the quadratic potential, being flatter, requires lower values of $C_{\phi}$ than the quartic model to achieve the same number of e-folds of inflation in the sub-planckian regime. Notice that, although we need somewhat large values of $C_{\phi}$ to obtain 40-60 e-folds of inflation, this allows inflation to occur at sub-planckian field values, which is not possible in standard inflation and is therefore a very attractive feature of warm inflation. 

For monomial potentials, we can also derive the following relation:
\begin{equation}
\frac{\eta_{\phi}}{1+Q} = \left(\frac{2(2r-1)}{r}\right)\left(\frac{\rho_R}{V}\right)\left(\frac{1+Q}{Q}\right)~.
\end{equation}
We can thus see that, when the radiation energy density becomes equal to the inflaton energy density, $\rho_R=\rho_{\phi}\approx V$, the slow-roll condition $\eta_{\phi}<1+Q$ has already been violated. This means, in particular, the breakdown of the slow-roll equation for the radiation energy density, in Eq.~(\ref{sreom}), as $\dot{\rho_R}$ becomes significant and radiation soon takes over. We wish to ultimately calculate the gravitino yield after inflation and this means that we need to evolve the full set of equations into the radiation era. To do this we must numerically solve the equations of motion (\ref{IE}) and (\ref{red}), which we will discuss in the next section.


\section{Gravitino production in warm inflation}\label{Gravprod}

\subsection{Particle Masses}\label{particlemasses}

\begin{figure}[htbp]
\centering
\includegraphics[scale=1]{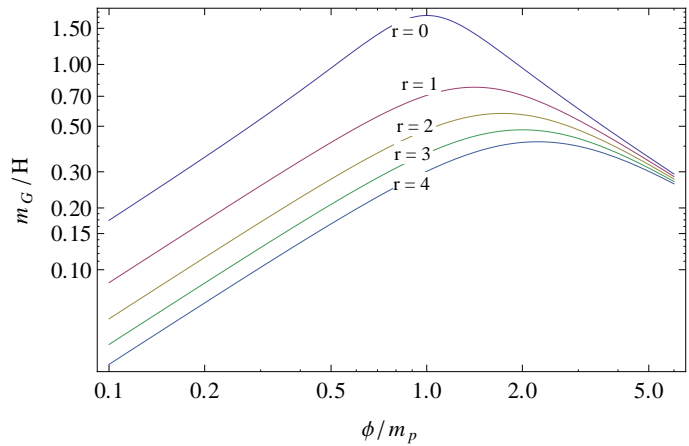}
\caption{$m_{\tilde{G}}/H$ for various monomial potentials.}
\label{Masses}
\end{figure}

In the presence of supersymmetry breaking, the gravitino gains a mass:
\begin{equation}
m_{\tilde{G}}=m_p\exp(-G/2)~.
\end{equation}
For the monomial superpotential in Eq.~(\ref{spinf}) and a canonical K\"ahler potential, the gravitino mass is then given by:
\begin{equation}
m_{\tilde{G}}=\frac{\lambda m_p}{r+1}\left(\frac{|\phi|}{m_p}\right)^{r+1}\exp\left(\frac{|\phi|^2}{2 m_p^2}\right)~.
\end{equation}
Comparing this to the Hubble parameter during inflation $H^2\approx(V/3m_p^2)$, we get
\begin{equation}
 \frac{m_{\tilde{G}}}{H} = \frac{\sqrt{3}}{r+1}\left(\frac{|\phi|}{m_p}\right)\left[1+\frac{1}{r+1}\left(\frac{|\phi|}{m_p}\right)^2\left(2-\frac{3}{r+1}\right)+\frac{1}{(r+1)^2}\left(\frac{|\phi|}{m_p}\right)^4\right]^{-1/2}~.
\end{equation}
As discussed above, we are interested in the sub-planckian regime, for which:
\begin{equation}
\frac{m_{\tilde{G}}}{H} \approx\frac{\sqrt{3}}{r+1}\left(\frac{|\phi|}{m_p}\right)~.
\end{equation}
In Figure \ref{Masses}, we can see that even if inflation is sub-planckian the gravitino mass can be a non-negligible fraction of the Hubble parameter during inflation, resulting in the gravitino mass being typically well above the TeV scale, in contrast with what was assumed in earlier works \cite{Bastero-gila}. For example, with a quartic potential, where $\lambda\sim10^{-7}$ yields the observed amplitude of density perturbations, if $\phi/m_p\approx1/2$ then $m_{\tilde{G}}\sim10^{10}\,$GeV.

As discussed earlier, thermal production of gravitinos procedes through the scattering of gauge bosons, gauginos, quark and squarks. It is, in particular, strongly dependent on the ratio of gaugino to gravitino masses, $m_{\tilde{g}}/m_{\tilde{G}}$. Having seen that supersymmetry breaking during inflation results in massive gravitinos, it is interesting to also consider its effect on the gaugino masses, which are given by the quadratic term in the Lagrangian:
\begin{equation}
\mathcal{L}_{{\text{gaugino}}}=\frac{m_p}{4}\exp(-G/2)G^l(G^{-1})^k_l\frac{\partial f^*_{\alpha\beta}}{\partial \phi^{*k}}\lambda^{\alpha}\lambda^{\beta}~,
\end{equation}
where $f_{\alpha\beta}$ is the gauge kinetic function, which is a holomorphic function of the chiral superfields in the model. It is dimensionless and symmetric with respect to its two adjoint indices and, in renormalisable theories, it is proportional to $\delta_{\alpha\beta}/g_{\alpha}^2$. Whether this function depends or not on the inflaton field is a model-dependent question and for completeness we will consider both cases separately. Interesting examples of inflaton-dependent gauge kinetic functions may arise in extra-dimensional theories such as superstring/M-theory, where the inflaton is identified with a modulus field (see e.g.~\cite{Ibanez:2012zz}). Considering the case where the inflaton's supersymmetry breaking effect is communicated to the visible sector through gravitational interactions, we can expand the gauge kinetic function in powers of $\phi/m_p$, yielding:
\begin{equation}
f_{\alpha\beta}\approx \delta_{\alpha\beta}\left(\frac{1}{g_{\alpha}^2}+f_{\alpha}\frac{\phi}{m_p}+...\right)~,
\end{equation}
where $f_{\alpha}$ is a dimensionless coupling, which for simplicity we will assume is universal to all the gauginos and will take to be $\mathcal{O}(1)$. Although the inflaton field modifies the gauge couplings, this will not change the running of the couplings significantly since are considering sub-planckian field values. With the above expansion for the gauge kinetic function, the gaugino mass is given by:
\begin{equation}
m_{\tilde{g}}= \frac{m_p}{4}\frac{\lambda}{r+1}\left(\frac{|\phi|}{m_p}\right)^{r+2}\left(1+(r+1)\left(\frac{\phi}{m_p}\right)^{-2}\right)f\exp\left(\frac{|\phi|^2}{2m_p^2}\right)~,
\end{equation}
and for sub-planckian field values this reduces to:
\begin{equation}
 m_{\tilde{g}} \approx \frac{\lambda m_p}{4}\left(\frac{|\phi|}{m_p}\right)^rf = \frac{\sqrt{V}}{4m_p}f = \frac{\sqrt{3}}{4}Hf~.
\end{equation}
The gaugino masses are thus proportional to the Hubble parameter. We then find that:
\begin{equation}
 \frac{m_{\tilde{g}}}{m_{\tilde{G}}}\approx\frac{(r+1)f}{4}\left(\frac{\phi}{m_p}\right)^{-1}~,
\end{equation}
so that gauginos are generically heavier than gravitinos during inflation, which will have important consequences on gravitino production. 

If the only source of supersymmetry breaking were the inflaton superpotential, then it is evident that as the inflaton rolls to its minimum supersymmetry would be restored. This is obviously not the case in nature, and so we will consider a supersymmetry breaking contribution from a hidden sector that gives rise to TeV-scale supersymmetric partners. The details of this hidden sector will not be important to the thermal production mechanism and so we can take the following phenomenological approximation for the masses:
\begin{eqnarray}
m_{\tilde{G}}&=&m_{\tilde{G}_{\phi}}+m_{\tilde{G}_0}~, \\
m_{\tilde{g}_i}&=& m_{\tilde{g}_{i \phi}}+m_{1/2}\frac{g_i(T)^2}{g(T_{{\text{GUT}}})^2}~,
\end{eqnarray}
where the subscript `$\phi$' denotes the inflaton contribution and `$0$' indicates the low-energy hidden sector contribution.


\subsection{Gravitino yield evolution}

We numerically solve the warm inflation equations (\ref{IE}) and (\ref{red}) along with the Boltzmann equation for the gravitino number density, Eq.~(\ref{BoltzGrav}), with the collision term given by Eq.~(\ref{Collision}). We run the couplings and gaugino masses with temperature at one-loop assuming they unify at the GUT scale, $T_{{\text{GUT}}}=2\times10^{16}\,$GeV, with universal gaugino mass $m_{1/2}=400\,$GeV. For convenience, we evolve the equations in terms of number of e-folds, $Hdt=dN_e$, which we will use both during and after inflation. We find that the thermally produced gravitino yield freezes out and approaches a constant value after inflation ends when the following three conditions are met:
\begin{itemize}
\item The gravitino has settled to its low-energy mass, given by the hidden sector contribution $m_{\tilde{G}_0}$;
\item The universe is in the radiation-dominated regime, where $\rho_R\sim \exp(-4N_e)$, i.e. radiation must have ceased being significantly produced by dissipation;
\item Thermal production of gravitinos must have stopped, so that the collision term in the Boltzmann equation is negligible and thus the number density evolves as $n_{\tilde{G}}\sim \exp(-3N_e)$.
\end{itemize}

A helpful consequence of being in the sub-planckian regime is that the large value of $C_{\phi}$ makes the primordial yield independent of $(T_*/H_*)$ in both quadratic and quartic models. 

\begin{figure}[htbp]
\centering
\subfigure{
\includegraphics[scale=0.9]{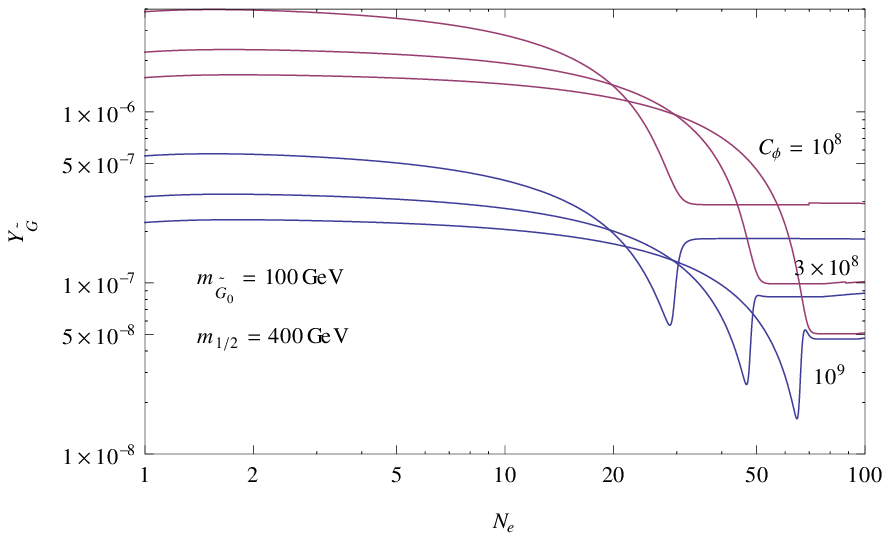} 
}
\subfigure{
\includegraphics[scale=0.9]{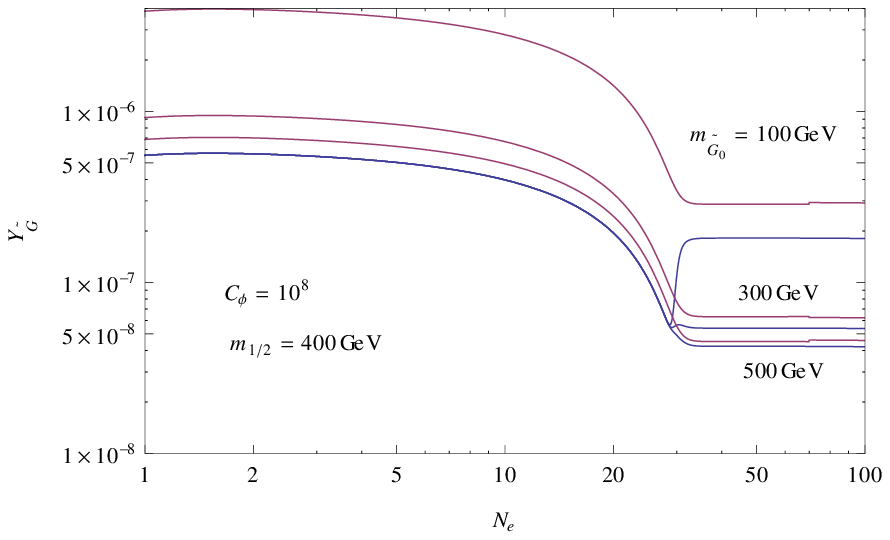}
}
\caption[Optional caption for list of figures]{The gravitino yield as a function of the number of e-folds for the quartic potential with inflaton-dependent (blue) and independent (red) gravitino mass, with inflaton-independent gaugino masses in both cases. In the left plot
we vary $C_{\phi}$ and in the right plot we vary $m_{\tilde{G}_0}$}
\label{Yieldplots}
\end{figure}

Having focused on the end of inflation, previous analyses have neglected the contribution from the non-vanishing inflaton value to the gravitino mass. To estimate the significance of this effect, we also consider the evolution of the gravitino yield for the unrealistic case where $m_{\tilde{G}}=m_{\tilde{G}_0}$ throughout inflation, and in Figure \ref{Yieldplots} we show an example of our results for a quartic potential in both cases, with inflaton-independent gaugino masses. During inflation the true yield is suppressed compared to the inflaton-independent gravitino yield due to the large gravitino mass supressing the collision term in Eq.~(\ref{Collision}). We then observe a sudden increase in the true yield as the gravitino mass rapidly decreases and settles to its low-energy value, causing the $m^2_{\tilde{g}}/{m^2_{\tilde{G}}}$ term to dominate. The yield increases until the gravitino mass reaches $m_{\tilde{G}_0}$ and it is then just a matter of a few e-folds until the collision term becomes negligible and radiation fully dominates, at which point the yield freezes out.

\begin{figure}[htbp]
\centering
\includegraphics[scale=1]{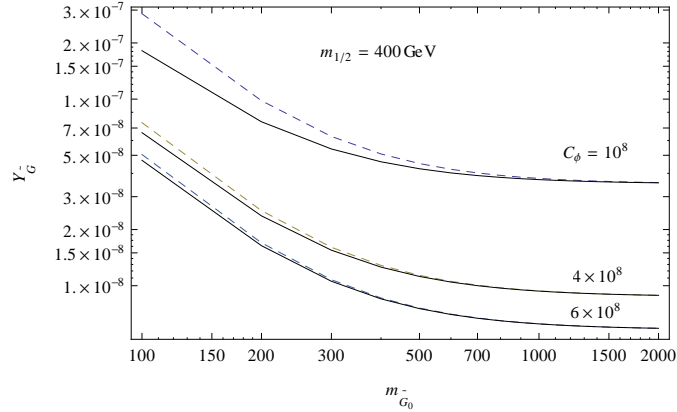}
\caption{Comparison between the thermally produced yield after freeze-out for inflaton-dependent (solid) and inflaton-independent (dashed) gravitino masses with a quartic potential and inflaton-independent gaugino masses.}
\label{constgmass}
\end{figure}

In Figure \ref{constgmass}, we show the difference in the thermal gravitino yield after freeze-out between inflaton-dependent and inflaton-independent gravitino masses. For large $C_{\phi}$ there is a negligible difference between the two cases. Increasing $m_{\tilde{G}_0}$ reduces this difference, due to the $m^2_{\tilde{g}}/{m^2_{\tilde{G}}}$ term never dominating, and in the large $m_{\tilde{G}_0}$ limit $C_{\tilde{G}}\sim T^6$.

\begin{figure}[htbp]
\centering
\includegraphics[scale=1]{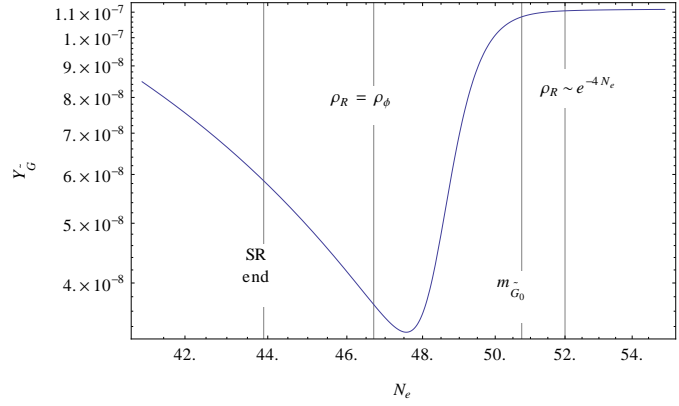}
\caption{The gravitino yield as a function of the number of e-folds for a quartic potential, indicating the number of e-folds at which slow-roll ends, $\rho_R\approx\rho_{\phi}$, $m_{\tilde{G}}\approx m_{\tilde{G}_0}$ and $\rho_R\sim\exp(-4N_e)$. These results correspond to $C_{\phi}=2\times10^8$, $(T_*/H_*)=1000$ and $m_{\tilde{G}_0}=100\,$GeV.}
\label{Yieldconstraints}
\end{figure}

In Figure \ref{Yieldconstraints}, we plot the gravitino yield as a function of the number of e-folds, indicating where the above conditions are met. It is clear from this figure that the yield is not yet constant when $\rho_R=\rho_{\phi}$ and that it changes quite drastically over a short number of e-folds until it becomes constant. Moreover, for the same parameters as in Figure \ref{Yieldconstraints}, applying the standard reheating constraints using the temperature at which $\rho_R=\rho_{\phi}$ or the temperature at which the yield freezes out results in $Y_{\tilde{G}}\sim10^{-9}$ and $Y_{\tilde{G}}\sim10^{-10}$, respectively, which are a few orders of magnitude lower than the true yield. This is due to the cumulative effect of gravitino production throughout warm inflation and that has been neglected in earlier analyses of this problem.

\begin{figure}[htbp]
\centering
\includegraphics[scale=1]{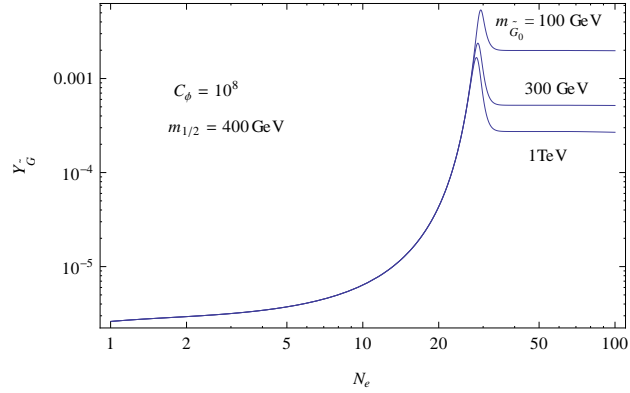}
\caption{The gravitino yield for inflaton-dependent gaugino masses as a function of the number of e-folds, for a quartic potential with $C_{\phi}=10^8$ and different values of $m_{\tilde{G}_0}$.}
\label{GMvarygrav}
\end{figure}

Figure \ref{GMvarygrav} shows the gravitino yield as a function of the number of e-folds for inflaton-dependent gaugino masses. We observe that, during inflation, gauginos are heavier than the gravitino and so the yield is larger than in the case where the gaugino masses do not depend on the inflaton field, once again due to the $m^2_{\tilde{g}}/{m^2_{\tilde{G}}}$ term in Eq.~(\ref{Collision}). The rise in $Y_{\tilde{G}}$ is due to $m_{\tilde{g}}/m_{\tilde{G}}\sim (\phi/m_p)^{-1}$, so that as the inflaton field decreases this term enhances the yield until the gravitino mass settles at its low-energy value, $m_{\tilde{G}_0}$. As before, it is only a matter of a few e-folds until the collision term becomes negligible and the universe is in the radiation era. During this short number of e-folds the collision term evolves as $C_{\tilde{G}}\sim T^6$ and so the yield decreases until it freezes out. We also find that, as expected, the lower the value of $m_{\tilde{G}_0}$, the larger the final yield is.

\subsection{Stable gravitinos}

\begin{figure}
\centering
\subfigure{
\includegraphics[scale=0.8]{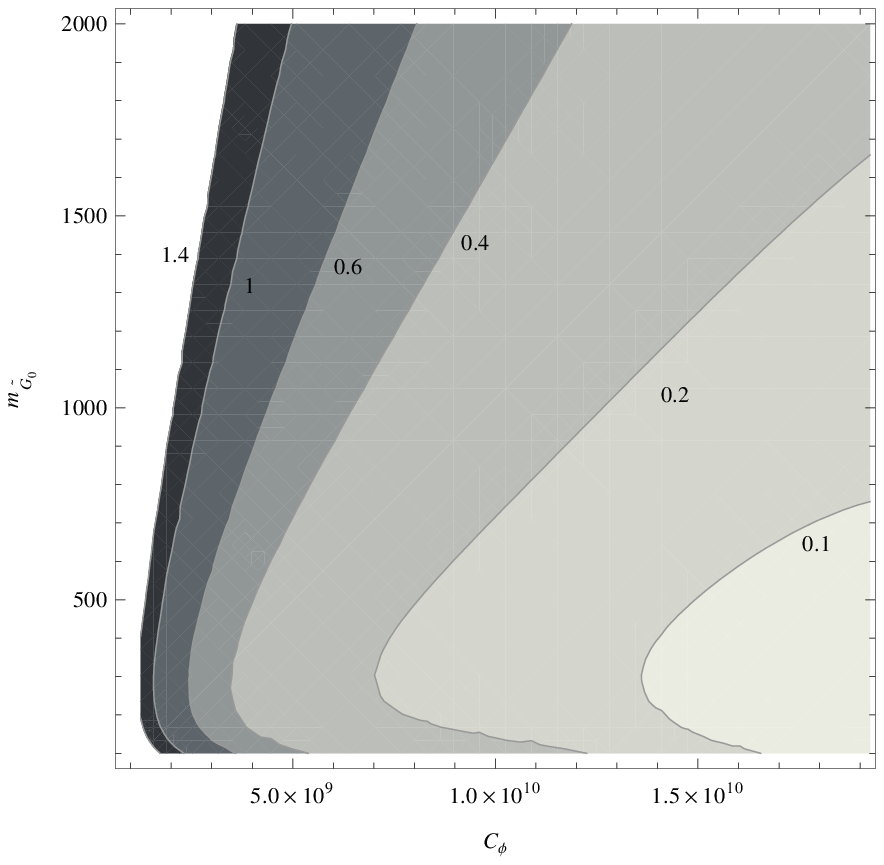} 
}
\subfigure{
\includegraphics[scale=0.8]{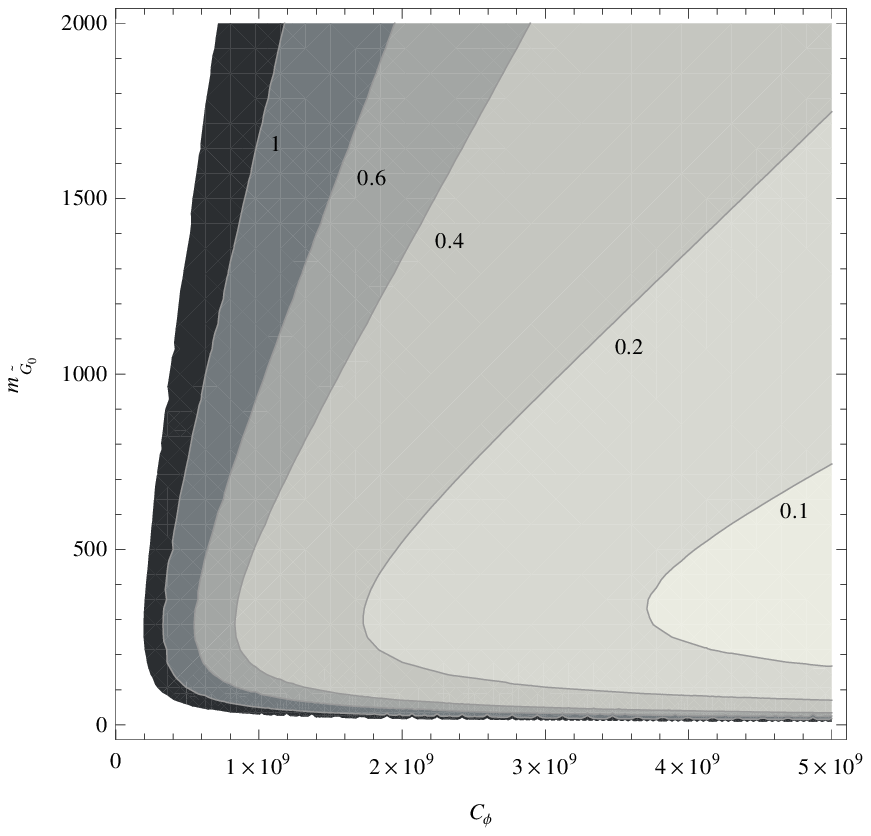}
}
\caption[Optional caption for list of figures]{Contribution to the density parameter $\Omega_{\tilde{G}} h^2$ from an LSP gravitino for the quartic (left) and the quadratic (right) potentials, with inflaton-independent gaugino masses. Masses are given in GeV.}
\label{Omega}
\end{figure}

\begin{figure}
\centering
\subfigure{
\includegraphics[scale=0.8]{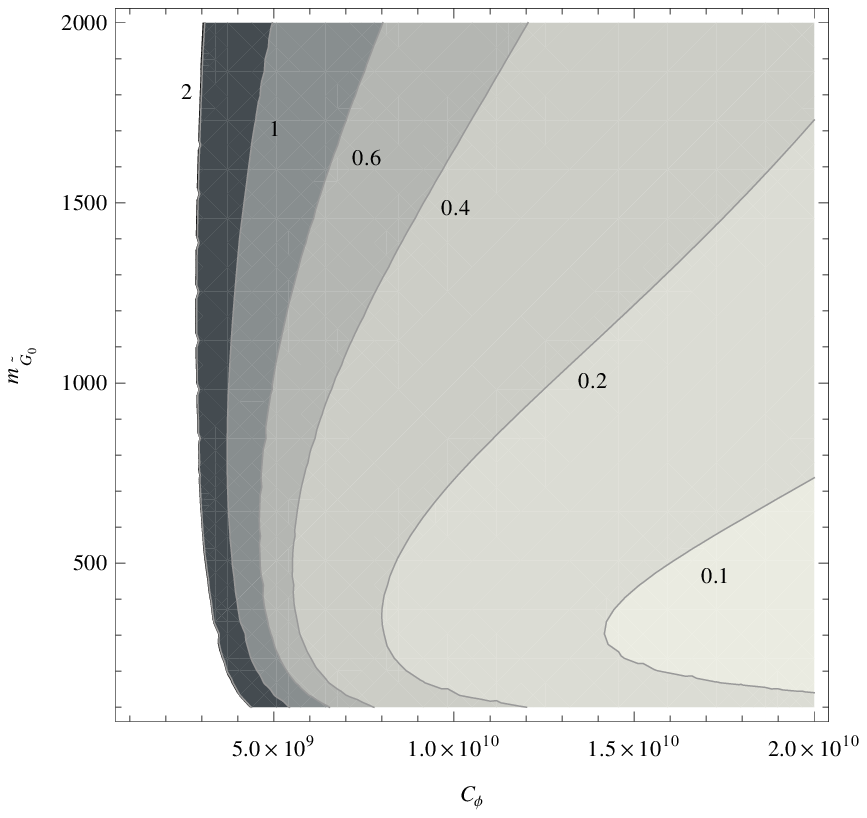} 
}
\subfigure{
\includegraphics[scale=0.8]{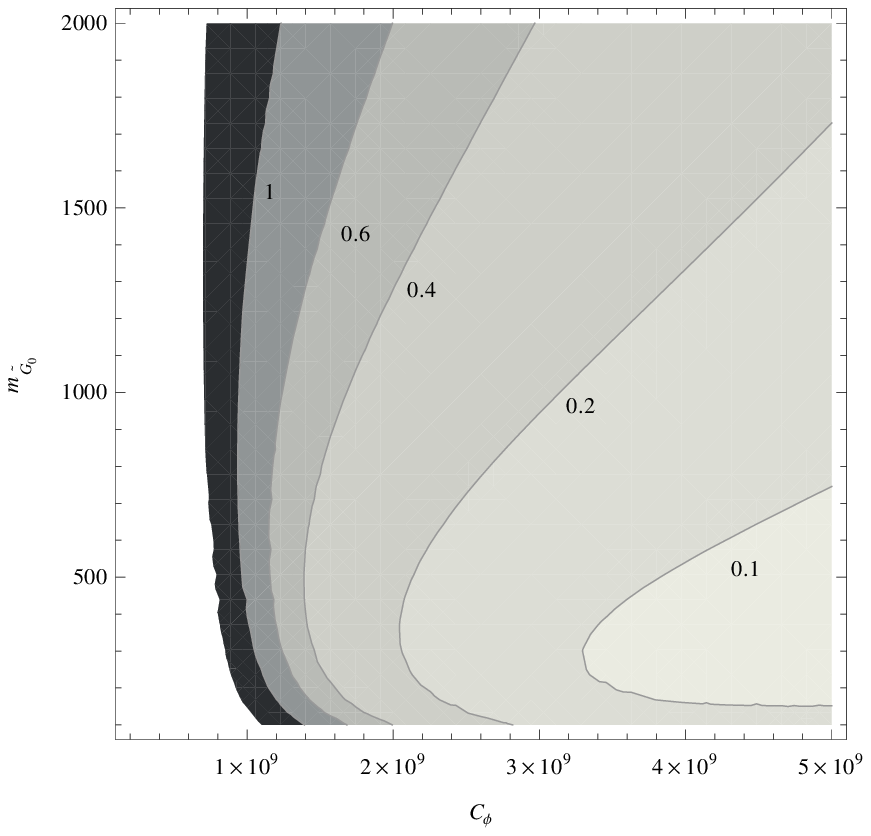}
}
\caption[Optional caption for list of figures]{Contribution to the density parameter $\Omega_{\tilde{G}} h^2$ from an LSP gravitino for the quartic (left) and the quadratic (right) potentials, with inflaton-dependent gaugino masses. Masses are given in GeV.}
\label{GMOmega}
\end{figure}

Figure \ref{Omega} shows the contribution of stable gravitinos to the current density parameter, $\Omega_{\tilde{G}} h^2$, for the quartic and quadratic potentials for inflaton-independent gaugino masses. Similarly, in Figure \ref{GMOmega} we plot this contribution for inflaton-dependent gaugino masses. 


We can see that for sufficiently large $C_{\phi}$, it is possible to avoid overclosure, $\Omega_{\tilde{G}} h^2\le 1$, for a broad range of gravitino masses. This is related to the fact that increasing $C_{\phi}$ reduces the temperature of the radiation bath during inflation and hence reduces the thermal production. This can be achieved with lower values of $C_{\phi}$ in the quadratic model than the quartic, due to the former being flatter. If the gaugino masses depend on the inflaton, the overclosure problem becomes more severe. We can nevertheless satisfy the dark matter constraint for LSP gravitinos, $\Omega_{\tilde{G}} h^2\lesssim0.1$, if $C_{\phi}\gtrsim 1.5\times10^{10}$ for the quartic and $C_{\phi}\gtrsim 4\times 10^9$ for the quadratic potentials. At these large values of $C_{\phi}$, there is little difference between inflaton-dependent and independent gaugino masses scenarios. 

\begin{figure}[htbp]
\centering
\subfigure{
\includegraphics[scale=0.9]{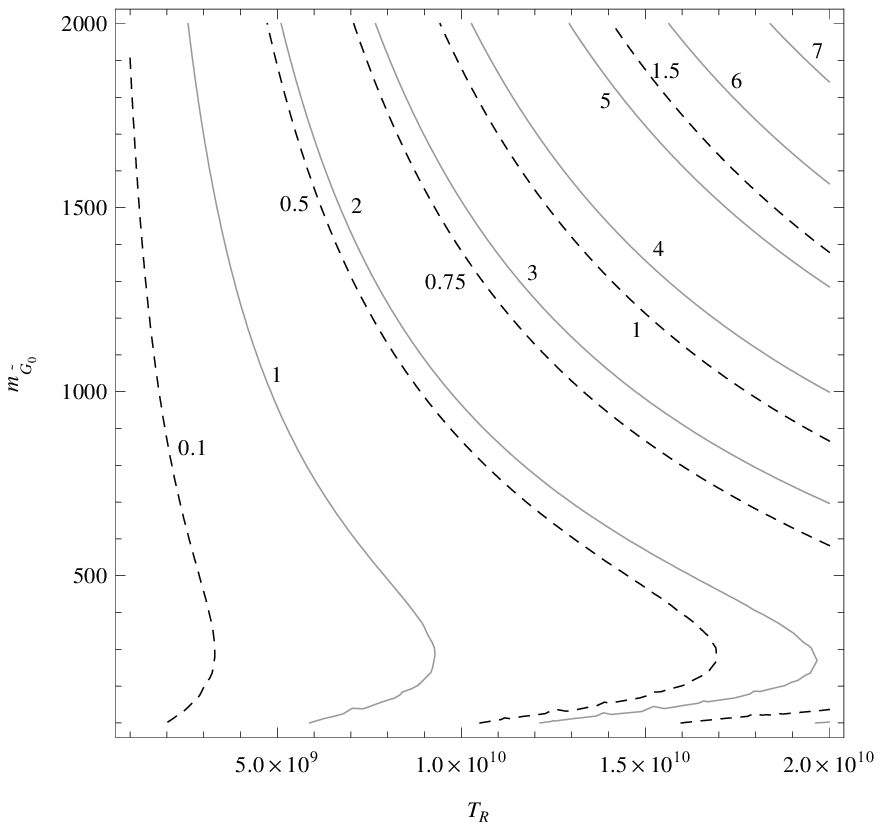} 
}
\subfigure{
\includegraphics[scale=0.9]{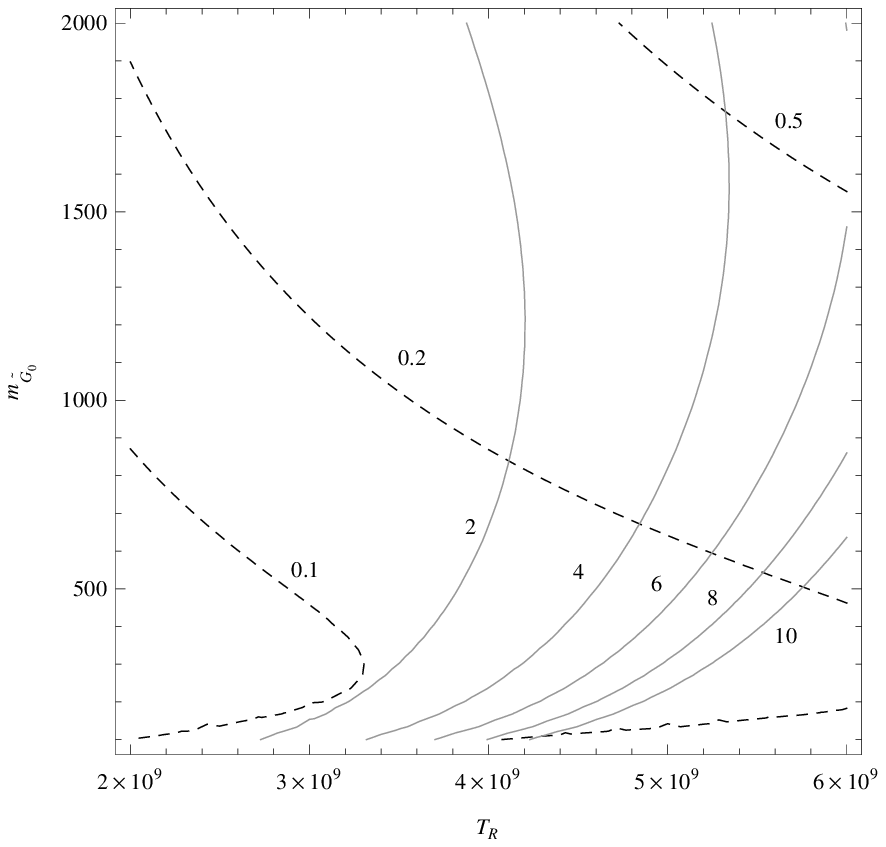}
}
\caption[Optional caption for list of figures]{Comparison between the predictions for $\Omega_{\tilde{G}} h^2$ using Eq.~(\ref{Steffan}) at the effective reheat temperature (dashed) and the full numerical simulation (solid) for inflaton-independent gaugino masses (left) and inflaton-dependent gaugino masses (right), with a quartic potential. All quantities are given in GeV.}
\label{OmegaCompare}
\end{figure}

For comparison with standard reheating predictions, we may define an effective reheat temperature as the temperature at which the gravitino yield becomes constant. In Figure \ref{OmegaCompare}, we illustrate the difference between the results predicted by Eq.~(\ref{Steffan}) at this effective temperature with those obtained with the full numerical simulation for a quartic potential. 

We can conclude that, if the gaugino masses depend on the inflaton field, the standard reheating prediction is drastically different from the true warm inflation result, where the gravitino problem is more severe. If the gaugino masses are inflaton-independent then the standard prediction also leads to an underestimation of $\Omega_{\tilde{G}}h^2$. This implies that in warm inflation the effective reheat temperature needs to be somewhat lower than in standard reheating in order to avoid overclosure. For example, for a $1\,$TeV gravitino, standard constraints require $T_R\lesssim2\times10^{10}\,$GeV for $\Omega_{\tilde{G}}h^2\lesssim1$, whereas in warm inflation we require $T_R\lesssim5\times10^{9}\,$GeV. 

\begin{figure}[htbp]
\centering
\includegraphics[scale=0.95]{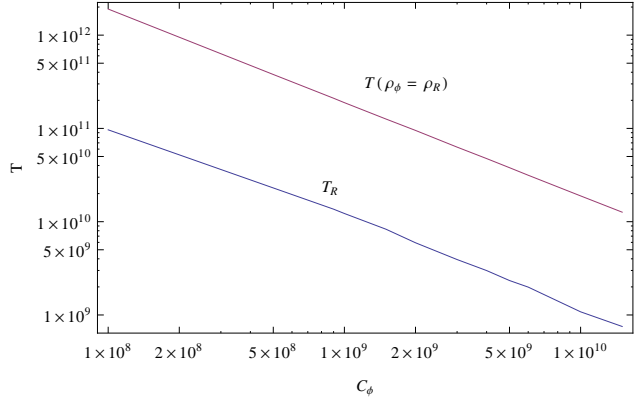}
\caption{The temperature, in GeV, at which $\rho_{\phi}=\rho_R$ and the effective reheat temperature, $T_R$, at which the gravitino yield freezes out, as a function of $C_{\phi}$ for quartic potential.}
\label{temperature}
\end{figure}

As discussed earlier, in the cold inflation picture it is assumed that the yield of gravitinos at the reheat temperature is zero, $Y_{\tilde{G}}(T_R)=0$ (see Section \ref{SGC}). This is perfectly valid in cold inflation, where due to the absence of a thermal bath during inflation, gravitinos are not produced. However, in warm inflation gravitinos are produced for the duration of inflation and so there is a non-negligible yield at the effective reheat temperature. Referring to Eq.(\ref{yieldevol}) and ignoring the temperature dependence of the masses and couplings, we see that in a Hubble time the gravitino yield behaves as $\Delta Y_{\tilde{G}}\sim T(\rho_R/\rho_{\phi})^{1/2}$. Even though $\rho_R/\rho_{\phi}$ is increasing during inflation, the temperature is decreasing and so, for monomial potentials in the strong dissipative regime, $\Delta Y_{\tilde{G}}\sim \phi^{2r/7}$, which decreases. The gravitino yield is thus non-negligible during warm inflation and in fact larger than the final value. Moreover, previous analyses of gravitino production during warm inflation assumed not only that the standard analysis at the end of inflation was applicable, but also that the gravitino yield froze out when $\rho_{\phi}=\rho_R$. We have seen that both these assumptions do not yield a good estimate for the gravitino abundance, both due to the cumulative effect of gravitino production during inflation and the fact that freeze-out does not occur until the universe if fully radiation-dominated, which occurs a few e-folds after inflaton-radiation equality. In particular, this results in an effective reheat temperature lower than previously estimated by more than one order of magnitude, as illustrated in Figure \ref{temperature}. 

Constraints on the LSP gravitino also come from decays of the NLSP spoiling BBN predictions for light-element abundances. It is typical to assume that the NLSP is the MSSM-LSP and that it will only decay into the gravitino and Standard Model particles. The NLSP lifetime typically depends upon the gravitino mass and their mass difference, $m_{{\text{NLSP}}}-m_{\tilde{G}}$, and so unless the gravitino is light and/or the NLSP is very heavy, it will be subject to BBN constraints. As the NLSP does not have planck-suppressed interactions with the other particles in the thermal bath, it may be in thermal equilibrium during inflation and freeze out in the radiation era. Its thermally produced yield will then be given by the freeze-out temperature, which places constraints on the NLSP and low scale gravitino masses but not on the warm inflation dynamics. In this respect the situation is the same as in cold inflation and, given that this is a model-dependent issue, we will not explore it any further, pointing the interested reader to the reviews in \cite{Feng2004,Feng2004a}.


\subsection{Unstable gravitino}

If the gravitino is the NLSP then we have the constraints from BBN on the primordial yield given in Section {\ref{SGC}, and in order to obtain such low yields we must consider large values of $C_{\phi}$. For $m_{\tilde{G}}\approx100\,$GeV the bound on the gravitino-to-entropy yield is $Y_{\tilde{G}}^s\lesssim10^{-16}$, which translates into $C_{\phi}\gtrsim10^{15}$. Similarly, for $m_{\tilde{G}}=1\,$TeV the bounds are $Y_{\tilde{G}}^s\lesssim10^{-14}$ and $Y_{\tilde{G}}^s\lesssim10^{-17}$ for branching ratios into hadrons of $B_h=10^{-3}$ and $B_h=1$, respectively. This requires $C_{\phi}\gtrsim10^{12}$ and $C_{\phi}\gtrsim10^{15}$, which are approximately the same for both the quartic and the quadratic potentials.

Also, in the case of a gravitino NLSP, each gravitino will then decay into one LSP. We can convert the primordial gravitino yield into the LSP yield using Eq.~(\ref{LSPconvert}). In Figure \ref{LSP}, we show the lines at which $\Omega_{LSP} h^2=0.1$ for a range of values of $C_{\phi}$ in the quartic and quadratic potentials.

\begin{figure}[ht]
\centering
\subfigure{
\includegraphics[scale=0.8]{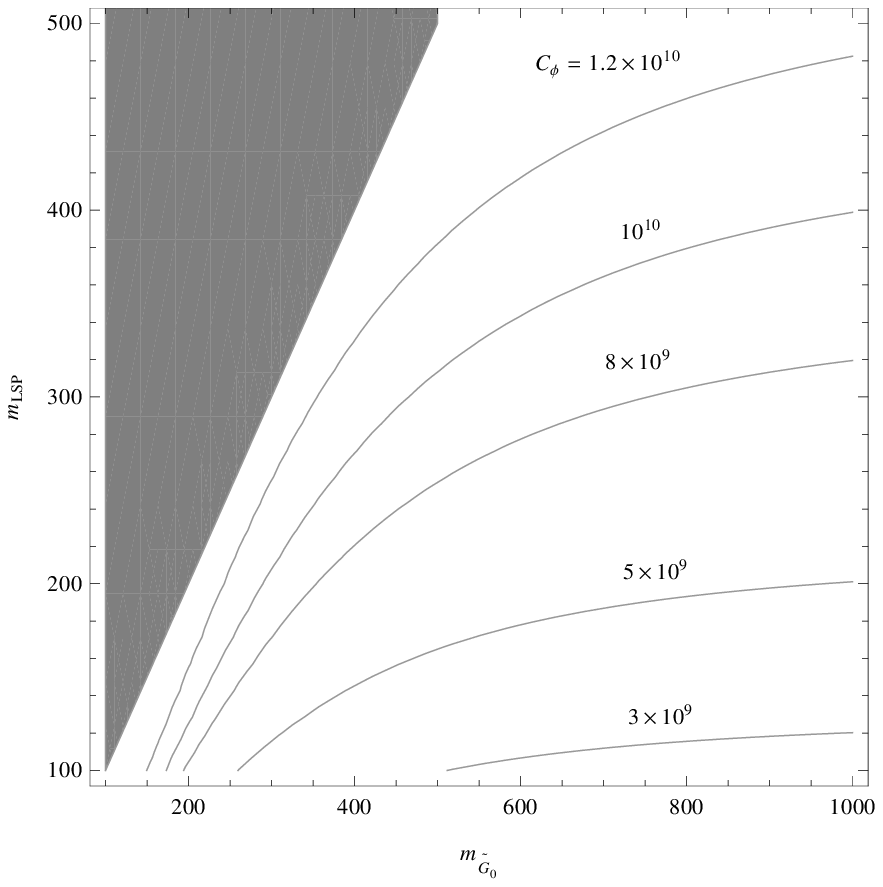} 
}
\subfigure{
\includegraphics[scale=0.8]{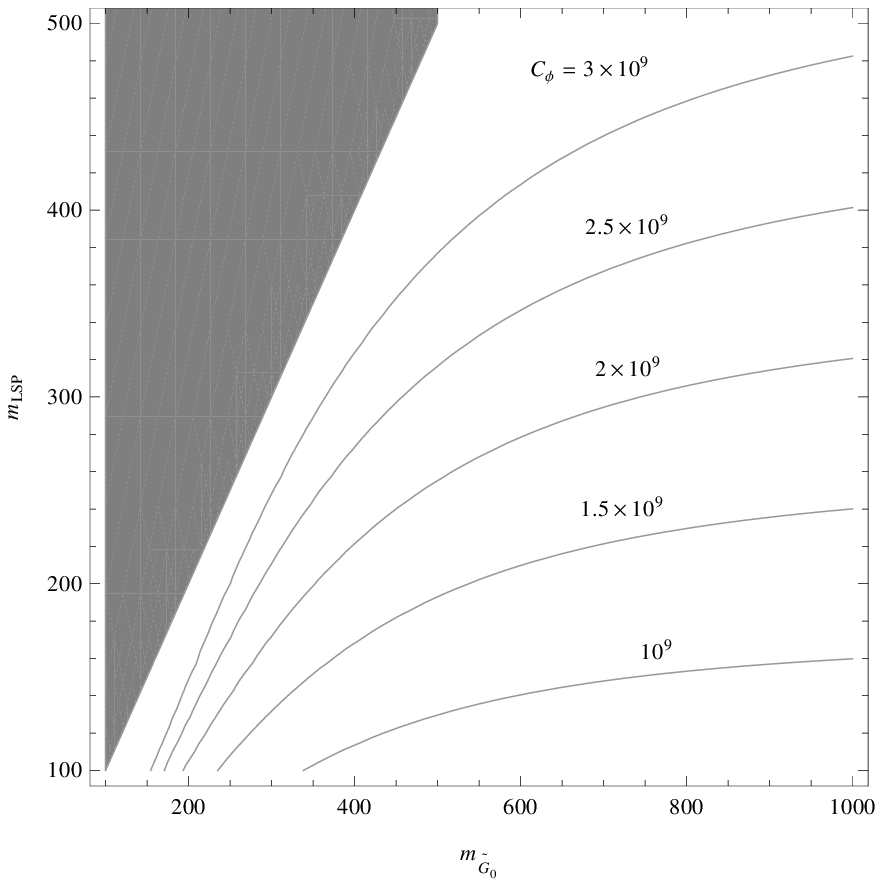}
}
\caption[Optional caption for list of figures]{Lines of $\Omega_{LSP} h^2=0.1$ for various values of $C_{\phi}$ in the quartic (left) and quadratic (right) potentials, with NLSP gravitinos. The shaded region indicates where the LSP is heavier than the gravitino. Masses are given in GeV.}
\label{LSP}
\end{figure}

We can see that the dark matter constraint can be satisfied for more reasonable values of $C_{\phi}$ than for the LSP gravitino. In particular, if $m_{{\text{LSP}}}=100\,$GeV and $m_{\tilde{G}_0}=1\,$TeV, the dark matter constraint is satisfied for $C_{\phi}\gtrsim 2.5\times10^9$ (quartic) and $C_{\phi}\gtrsim6\times10^8$ (quadratic). If the NLSP gravitino mass, $m_{\tilde{G}}\gtrsim20\,$TeV, then it decays before BBN and the strongest constraint is given by the dark matter bound on the LSP.  For $m_{\tilde{G}_0}=20\,$TeV the dark matter constraint is satisfied for $m_{{\text{LSP}}}=100\,$GeV with $C_{\phi}\gtrsim2.2\times10^{9}$ ($C_{\phi}\gtrsim6\times10^8$) and for $m_{{\text{LSP}}}=1\,$TeV with $C_{\phi}\gtrsim2.5\times10^{10}$ ($C_{\phi}\gtrsim6\times10^9$) in the quartic (quadratic) potential.


\section{Conclusion}\label{Discussion}

In this work, we have revisited the gravitino problem in warm inflation, focusing on thermal production which, providing the main difference from standard or cold inflation, places the strongest constraints on warm inflation dynamics. By performing a full numerical evolution of the gravitino yield into the radiation era we improve upon previous analyses. Firstly,  in the context of thermal gravitino production, the effective reheat temperature is the temperature at which the gravitino yield freezes out and not the temperature at which the inflaton energy density equals the radiation energy density. This allows the temperature to drop by approximately an order of magnitude, which lowers the final temperature at which gravitinos are produced compared to previous estimates. Secondly, we found that an analysis similar to standard reheating is in fact inadequate in describing gravitino production, due to the non-negligible yield produced throughout the whole duration of warm inflation. Finally, we have also taken into account the enhance particle masses during inflation due to supersymmetry breaking, in particular the gravitino and potentially the Standard Model gauginos. 

Taking all of these issues into account, our work shows, in particular, that the final gravitino yield is substantially lowered for stronger dissipative effects, as in practice this lowers the temperature of the radiation bath during warm inflation significantly. We have presented regions of parameter space where the LSP gravitino can satisfy the dark matter bound and, for an NLSP gravitino, we determined the regions where the LSP abundance does not exceed the amount of dark matter present in our universe  and have given values of the dissipation parameter $C_{\phi}$ for which late decays do not spoil the predictions of BBN.

Although thermal production is the dominant source of gravitinos during warm inflation, other non-thermal mechanisms may play a role at a later stage. Gravitinos can, in particular, also be produced from particle decays, but due to the large Hubble parameter during warm inflation these decays will not take place until the radiation era, at which point the standard cosmological results can be used. They can also be produced from the direct decay of the inflaton field, although we have found that, in the sub-planckian regime, the dissipative ratio $Q$ is necessarily large, which prevents the inflaton field from entering an oscillating phase. Figure \ref{Oscillations} shows the inflaton field evolution as we artificially switch off dissipation at $\rho_R=\rho_{\phi}$, at which point oscillations immediately begin. 

\begin{figure}[htbp]
\centering
\includegraphics[scale=1]{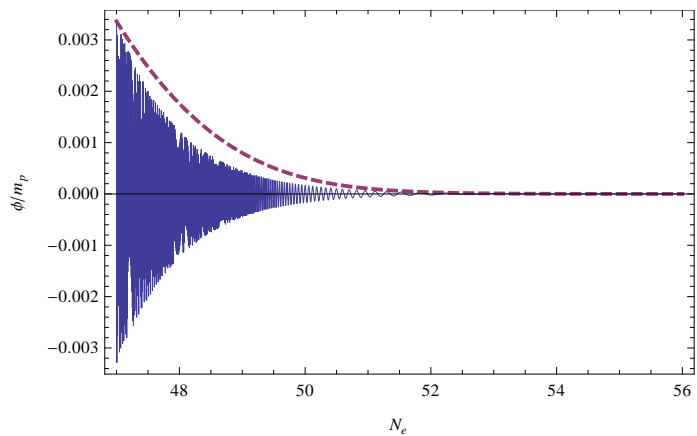}
\caption{Switching off dissipation at $\rho_R=\rho_{\phi}$, showing that the large dissipation keeps the inflaton field from oscillating in the radiation era. The dashed (solid) line corresponds to the case with (without) dissipation.}
\label{Oscillations}
\end{figure}

Dissipation will actually switch off when the heavy fields are no longer kinematically allowed to decay into the light degrees of freedom, which depends on their low scale mass hierarchy. For example, if supersymmetry is indeed a solution to the gauge hierarchy problem, we may expect light scalar masses to lie close to the TeV scale and dissipation to switch off at temperatures of this order. Apart from these kinematical constraints, the form of the dissipation coefficient in Eq.~(\ref{Upsilon}) may actually hold down to very low temperatures. For example, to avoid exceeding the dark matter bound for the LSP gravitino, we require $C_{\phi}\sim1.5\times10^{10}$ and for 40 e-folds of inflation, if the coupling $g\sim 1$, the system remains in the low-temperature regime down to $T\sim10\,$MeV, at which point $\rho_R/\rho_{\phi}\sim10^{12}$. It is therefore unlikely in this case that any oscillations of the inflaton field may come to play a significant role in gravitino or, in fact, any entropy production. In particular, a significant dilution of the gravitino yield through a late inflaton decay along the lines proposed in \cite{Bastero-gila} may be difficult to attain, although this may depend on the form of the inflaton potential, which goes beyond the scope of this work.

Our analysis revealed that it is possible to satisfy the dark matter constraint for LSP gravitinos and LSPs produced from NLSP gravitinos at large values of the dissipation parameter $C_{\phi}$, which requires large couplings and field multiplicites, pointing towards beyond the Standard Model scenarios. The gravitino problem is more severe for unstable gravitinos potentially spoiling the predictions of BBN, and in this case much larger values of $C_\phi$ are required. 

One should note that such large values of the dissipation coefficient are nevertheless required in order to overcome the severe eta-problem affecting monomial potentials for sub-planckian values. Above the Planck scale, the potential gets exponentially steeper with increasing field values, requiring larger values of $C_{\phi}$ to obtain 40-60 e-folds of inflation and also to suppress the resulting gravitino abundance. Using the full supergravity potential in Eq.~(\ref{scalar_potential}) places a lower bound of $C_{\phi}\gtrsim10^8$ and $C_{\phi}\gtrsim2\times10^7$ for 40 e-folds of inflation in the quartic and quadratic potentials, corresponding to $\phi_*\sim m_p$. Of course a non-canonical choice for the K\"ahler potential may alleviate this eta-problem, but supergravity is in any case unlikely to be the complete theory near the Planck scale and so any analysis along these lines must be taken with a pinch of salt. It should nevertheless be emphasized that simple monomial potentials cannot yield the required number of e-folds for sub-planckian values without dissipation, which is an attractive feature of warm inflation despite the large field multiplicities and/or couplings required.

One should bear in mind that observations may pose some constraints on the amount of dissipation present when the relevant CMB scales exit the horizon during inflation. In particular, an earlier analysis of non-gaussian effects on the primordial power spectrum showed that these depend logarithmically on the dissipative ratio at horizon crossing, $Q_*$, placing a model-dependent upper bound on the parameter $C_\phi$ \cite{Moss2007}. This analysis assumed, however, a constant dissipation coefficient, and more recently it was shown that for a generic $T$-dependence the non-gaussian parameter $f_{NL}$ is largely independent of the value of $Q_*$ in the strong dissipative regime, yielding $f_{NL}\sim\mathcal{O}(10)$ within the observable window of Planck \cite{Moss:2011qc}. Hence, although the dynamics of second-order perturbations in warm inflation is not yet fully understood, we do not expect non-gaussianity to pose any significant constraints on our results.

In this work, we have considered a general scenario where all the MSSM degrees of freedom are in thermal equilibrium during inflation. However, it has been pointed out in \cite{Rosa2012} that, in the low-temperature regime, $m_X\ll T$, fermionic degrees of freedom may actually not thermalize, as both their contribution to the dissipation coefficient and their thermal scattering cross section are supressed compared to scalar fields. This is related to the structure of the superpotential (\ref{twostage}) and the broken supersymmetry during inflation, which imply that the light fermions in the $Y$ multiplets only interact via the heavy $X$ bosons and fermions, whereas the light scalars have unsupressed interactions. Moreover, although the effects of gauge fields and their superpartners on the dissipation coefficient have yet to be analyzed in detail, their contributions to the dissipation coefficient may also be suppressed for sufficiently small gauge couplings. This would imply a thermal bath concurrent with inflation essentially composed of scalar particles, which would prevent gravitino production during inflation and eliminate the cumulative effect observed in our numerical simulations, at the same time requiring somewhat lower values of $C_\phi$ for sub-planckian inflation. Both fermionic and gauge degrees of freedom will nevertheless be `reheaten' after inflation with either the exit from the low-temperature regime or the Hubble parameter dropping sufficiently in the radiation era. Although it requires further investigation, this may occur only at very low temperatures, as discussed above, in which case thermal gravitino production will be negligible.

In cold inflation there is a tension between having a large enough reheat temperature for thermal baryogenesis/leptogenesis to occur and it being low enough to avoid overproduction of gravitinos and other unwanted relics (see e.g. \cite{Cline:2006ts}). In warm inflation this can be alleviated, as a baryon asymmetry can be produced through dissipation itself \cite{Bastero-Gil2011}. Dissipation is an inherently out-of-equilibrium process, so the inclusion of baryon number and CP-violating interactions in the $X$ and $Y$ sectors in the superpotential Eq.~(\ref{twostage}) naturally leads to the production of a baryon asymmetry during inflation. In the low-temperature regime, the produced asymmetry is naturally small despite the large couplings and field multiplicities required for a sufficiently long period of accelerated expansion and, moreover, this may lead to distinctive baryon isocurvature perturbations in the CMB anisotropies spectrum that may be observable in the near future. Warm inflation thus exhibits several attractive features that address not only the problems of inflationary dynamics itself but also many of the associated cosmological puzzles.

We would like to point out that the results from our analysis have a certain amount of crossover with cold inflation. Standard reheating is unlikely to be instantaneous and so the production of gravitinos will occur for the duration of the reheating phase. This will lead to an accumulated abundance similar to the one we have observed in warm inflation and so may change the standard reheating temperature constraints. The gravitino gets a mass from inflation and so, when the inflaton is oscillating about its minimum, the gravitino mass will also change at the same rate. If the oscillations are adiabatic, $\dot{m}_{\tilde{G}}/m_{\tilde{G}} \lesssim \Gamma_{{\text{scattering}}}$, then this effect can be analysed for various potentials in a similar way to the analysis performed in this work. It may then result in significant differences in the thermal production of gravitinos during the standard reheating picture.

With this work, we hope to have shed some light on gravitino production in warm inflation, with the way now paved for other potentials and dissipative coefficients to be analysed. In particular, the fact that inflation gives a mass to the gravitino may have a more significant impact on the thermal production in other potentials. Our analysis also brought to light some issues that may be significant to standard reheating and we hope that this motivates further exploration of this topic.


\begin{acknowledgments}
 This work was supported by the Science and Technology Facilities Council (United Kingdom).
\end{acknowledgments}


\bibliographystyle{apsrev} 
                 
\bibliography{Gravitinopaper.bib}

\end{document}